# Visualizing the interplay of Dirac mass gap and magnetism at nanoscale in intrinsic magnetic topological insulators

Mengke Liu[1], Chao Lei[1], Hyunsue Kim[1], Yanxing Li[1], Lisa Frammolino[1], Jiaqiang Yan[2], Allan H. Macdonald[1]*, Chih-Kang Shih[1]*

[1] Department of Physics, The University of Texas at Austin, Austin, TX 78712, USA

[2] Materials Science and Technology Division, Oak Ridge National Laboratory, Oak Ridge, TN 37831, USA

* To whom correspondence may be addressed.

**Email:** macdpc@physics.utexas.edu or shih@physics.utexas.edu.

**Abstract**

In intrinsic magnetic topological insulators, Dirac surface state gaps are prerequisites for quantum anomalous Hall and axion insulating states. Unambiguous experimental identification of these gaps has proved to be a challenge, however. Here we use molecular beam epitaxy to grow intrinsic $MnBi_2Te_4$ thin films. Using scanning tunneling microscopy/spectroscopy, we directly visualize the Dirac mass gap and its disappearance below and above the magnetic order temperature. We further reveal the interplay of Dirac mass gaps and local magnetic defects. We find that in high defect regions, the Dirac mass gap collapses. *Ab initio* and coupled Dirac cone model calculations provide insight into the microscopic origin of the correlation between defect density and spatial gap variations. This work provides unambiguous identification of the Dirac mass gap in $MnBi_2Te_4$, and by revealing the microscopic origin of its gap variation, establishes a material design principle for realizing exotic states in intrinsic magnetic topological insulators.

**Main Text**

**Introduction**

In magnetic topological insulators (MTI), time-reversal symmetry breaking opens a surface state Dirac mass gap that is a prerequisite for realizing quantum anomalous Hall (QAH) and axion insulator ground states (1-3). QAH states support spin-momentum locked dissipationless edge states that hold great promise in advancing electronic and spintronic device applications. Early materials design was based on doping topological insulators with magnetic elements (4, 5); however, dopant disorder (6, 7) is significant and considered to be the main cause for quantized edge states being observed only at extremely low temperature, ~ 30 mK in chromium-doped $(Bi,Sb)_2Te_3$ (5). Recently a new family of intrinsic MTIs based on layered van der Waals material $MnBi_2Te_4$ has emerged, in which an ordered Mn magnetic layer is incorporated stoichiometrically in the middle of each van der Waals layer (8-14), suppressing the disorder effect. Indeed, the QAH effect has been observed in $MnBi_2Te_4$ at a significantly higher temperature of 1.4 K (15). However, the QAH effect is not always observed, even in samples that appear similar (16-20). Moreover, observations of the Dirac mass gap, a prerequisite for the QAH effect and a direct consequence of exchange coupling between the Dirac surface state and the spontaneous magnetization (21, 22), has proven to be challenging and remained unclear (9, 23-30). Given the significant fundamental and application interest in this system, addressing the elusive nature of the Dirac mass gap becomes ever more critical.

Current bulk $MnBi_2Te_4$ samples are known to be heavily degenerate n-type, with Fermi level ($E_F$) located deeply inside the conduction band (9, 23-26). As a result, short Dirac fermion lifetimes smear gap signatures. Besides, defects distribution inhomogeneity can complicate the magnetic order (30-32) and impact the exchange gap distribution (6, 32-37). Spatial fluctuations of chemical potential and another chemical composition mixing, such as $Bi_2Te_3$ and $MnTe_2$ (16, 17, 38), further entangle the electronic structure and complicate the gap observation. Thus, to properly address the nature of Dirac mass gap, a microscopic characterization based on intrinsic samples with Fermi level lies within the Dirac gap is crucial.

Here we achieve successful layer-by-layer growth of $MnBi_2Te_4$ thin film on HOPG and graphene substrate using molecular beam epitaxy (MBE). Furthermore, we control the defect concentration such that $E_F$ lies inside the Dirac gap. Using *in situ* scanning tunneling microscopy/spectroscopy (STM/S), we directly observed the Dirac mass gap and its variation as a function of film thickness and the magnetic antistites concentrations. We found that in high defect regions, the Dirac mass gap collapses. To understand these phenomena, we combine *ab initio* supercell calculation with the coupled Dirac cone model to account for the effect of magnetic antistites defects and successfully demonstrate the Dirac mass gap fragility to antisites. Our calculation predicts maximum antisite densities for observing the QAH effect. This work provides unambiguous evidence for the presence of a Dirac mass gap in intrinsic MTI and reveals the microscopic origin for the gap variation. It further establishes magnetic defect as a critical tuning knob in controlling the topological phases of MTI for QAH and axion insulator-based device applications.

**Bulk crystals versus MBE grown thin films**

$MnBi_2Te_4$ is a layered van der Waals material consisting of Te-Bi-Te-Mn-Te-Bi-Te septuple layers (SLs), in which the Mn atomic layer is coupled ferromagnetically within the SL and antiferromagnetically with neighboring SLs. The lattice illustration shown in Fig. 1A includes two commonly observed native defects, $Mn_{Bi}$ and $Bi_{Te}$ antisites. STM images taken on the cleaved bulk crystals (Fig. 1B) show ~5% $Mn_{Bi}$ antisites, which appear as dark triangular depressions (39), and ~0.2% of $Bi_{Te}$ antisties which appear as bright circular protrusions (39). The abundance of the $Bi_{Te}$ antisties is related to the Te content during the growth (40). The broad background fluctuation may reflect an inhomogeneous distribution of the $Bi_{Mn}$ antisite, the most common n-type dopant (40), located at the Mn layer and is too far from the surface to be directly imaged by STM. But its high concentration is expected from the internal strain energy in the middle Mn layer [40] and can be inferred from the heavily degenerate n-type nature of the sample. As shown in Fig. 1C, we find that $E_F$ is 0.28 eV above the Dirac point in bulk samples, consistent with previous ARPES results (9, 23-25). However, the Dirac gap signature is smeared and difficult to quantify.

We succeed in growing $MnBi_2Te_4$ thin films on the HOPG and the graphene substrates. The STM/S results reported here are primarily carried out on samples grown on HOPG substrate. A typical STM image of our MBE-grown $MnBi_2Te_4$ flakes, shown in Fig. 1D, demonstrates SL-by-SL growth of high-quality $MnBi_2Te_4$ thin film. A topographic section across the flakes (Fig. 1E) allows us to determine the thickness in terms of the number of SLs. This thickness determination is crucial because it eliminates the possibility of $Bi_2Se_3$ or $MnTe_2$ intermixing, which would lead to different layer thicknesses. (Additional topographic images of MBE grown samples can be found in Supplementary Fig. S1.) Fig. 1D inset shows the corresponding atomic image, exhibiting ~4% of $Mn_{Bi}$ antisites and no bright $Bi_{Te}$ antisites. The disappearance of $Bi_{Te}$ antisites is due to a Te-rich growth ambient which also reduces the $Bi_{Mn}$ antisites (40). Most importantly, in the ultra-thin regime, the internal energy in the Mn layer may not yet build up significantly, thus delaying the strain compensation effect through $Bi_{Mn}$ antisites (40). The clean and flat background of the atomic image reflects a considerably reduced concentration of $Bi_{Mn}$ antisites. The SL-by-SL MBE growth of MnBi$_2$Te$_4$ films with low defect density we report provides access to the rich electronic properties of MnBi$_2$Te$_4$ thin films.

**Direct observation of Dirac mass gaps**

We start with the 4 SL film, thick enough that its topological surface state is well-developed. Fig. 2A shows a typical tunneling spectrum acquired at 4.3 K at a relatively high setpoint bias of 0.6 V. In this spectrum, there exists a sizeable low-conductance bias range from -0.3 V to +0.4 V, marked by the blue and orange arrows. This low conductance region is due to a significantly reduced density of state (DOS). We find that with a lower setpoint bias as shown in Fig. 2B for which the tip-to-sample distance is reduced, states near $E_F$ can be well resolved. The behavior of bias-dependent tip-to-sample distance can be found in Supplementary Fig. S2, and the tunneling junction characteristics (barrier height) at different biases can be found in Supplementary Fig. S3. In Fig. 2B, which focuses on states near the Fermi energy, one can observe a clear exchange gap of ~ 120 meV with $E_F$ located 20 meV below the conduction band edge.

We further carried out the STS measurement at 77 K, a temperature above the Neel temperature. At a high setpoint bias (Fig. 2C), the spectrum exhibits a low conductance region from -0.3 V to 0.4 V, similar to that acquired in Fig. 2A. With a low setpoint bias (Fig. 2D), the electronic states near the Fermi level can be clearly observed. However, the gap disappears. This direct comparison further affirms that the observed gap at 4.3 K is the Dirac mass gap due to the exchange interaction of the Dirac surface states with the spontaneous magnetization below the Neel temperature.

Fig. 2E shows theoretical projected density-of-states (DOSs) curves for the top four atomic layers of a defect-free 4 SL MnBi$_2$Te$_4$ thin film. These results show a ~ 0.45 eV low DOS energy range (marked between the blue and orange arrows), in agreement with our experimental observed low conductance region of 0.7 eV. The true theoretical Dirac mass gap within which the DOS vanishes is only 48 meV for a 4 SL film. Fig. 2F summarizes the STS-measured (See Supplementary Fig. S4 for 1 to 3 SL STS) and the density-functional theory (DFT)-calculated (Supplementary Fig. S5) energy gap as a function of the thickness, which shows excellent overall agreement, given the known tendency toward bandgap under-estimation in DFT. Our gap determinations have been cross-checked by obtaining STS data at different temperatures and in several different measurement modes (Fig. S2 and S3). This provides unambiguous evidence of the Dirac mass gap in intrinsic magnetic topological insulators, resolving current debates (9, 21, 23-28).

**Role of magnetic antisite defects**

The thickness-dependent STS measurements discussed above were carried out in low defect areas, where STS shows consistent uniform gap values. We next discuss how defects impact the local electronic structure. Fig. 3A shows a region with low $Mn_{Bi}$ antisites (< 4%) on 4 SL $MnBi_2Te_4$. Fig. 3B displays spatial dependent tunneling spectra. Depending on their specific locations, the tunneling peak features show intensity variation. However, the spectra exhibit a uniform gap value with $E_F$ located within the gap. Note that the intrinsic nature of this sample suggests a similar amount of $Bi_{Mn}$ antisite (n-type dopant) and $Mn_{Bi}$ antisite (p-type dopant). This situation changes drastically in high defective areas. Fig. 3C shows the atomic image of the defective area on 3 SL $MnBi_2Te_4$, exhibiting high concentrations of $Mn_{Bi}$ (~ 9%). In such areas, locally "defect-free" regions on the scale of 2 nm by 4 nm can be found. Fig. 3D shows a sequence of spectra from points between the locally defect-free region (location #1) and the defective area. The spectrum at position #1 exhibits a finite Dirac gap, whose tunneling features are similar to those on the clean area (Supplementary Fig. S4C). As one progresses to a defective area, the Dirac gap decreases (position #2) and eventually collapses (#3 to #6) with a V-shape DOS suggesting the recovery of the linear Dirac dispersion. Interestingly, in the spectra acquired in the defective region (#3 to #6), one observes an additional zero-bias anomaly, characterized by a dip at zero bias and two peaks at $\pm 6\ meV$. We attribute this zero-bias anomaly to inelastic scattering due to spin excitations, similar to those reported previously in other systems (41-43), further confirming the magnetic nature of the $Mn_{Bi}$ antisites. The evolution of the Dirac mass gap observed here strongly suggests the significance of defects concentration, particularly magnetic antisites, in suppressing the exchanging coupling between the Dirac surface states and the magnetic moments.

### *Ab initio* DFT and coupled Dirac cone model calculations

Motivated by our experimental observations, we performed DFT supercell calculation to account for the role of defects and interpreted them using a simplified Dirac-cone model (22) of the $MnBi_2Te_4$ electronic structure. Fig. 4A illustrates the magnetic moments and exchange couplings present in our model calculation. The $Mn_{Bi}$ antisites moments have been shown to be antiparallel to those in the central Mn layer (31). Exchange interactions between Fermi level electrons and these moments partially cancel those from the central Mn layer, reducing the same-septuple-layer and neighboring-septuple-layer exchange couplings $J_s$ and $J_D$ (See Supplementary Fig. S6 for detailed definition.). To estimate the change of $J_s$ and $J_D$ as a function of magnetic antisites density, we perform DFT supercell calculations for bulk $MnBi_2Te_4$ crystals that have a ferromagnetic spin configuration and varying antisite defect configurations. A $4 \times 4 \times 1$ superstructure with one $Mn_{Bi}$ antisite per sixteen Bi atoms is used to simulate 6% $Mn_{Bi}$ (see Supplementary Fig. S7 for other superstructure configurations). To keep the chemical potential within the gap, we choose to compensate the $Mn_{Bi}$ antisite with a $Bi_{Mn}$ antisite by substituting the Bi atom with its next nearest-neighbor Mn atom, as illustrated in Fig. 4A. This antisite pair configuration has the lowest formation energy (31). The presence of $Bi_{Mn}$ antisites also decreases $J_S$ by reducing the magnetic moment in the Mn layer. As shown in Fig. 4B, bulk $MnBi_2Te_4$ with a ferromagnetic spin configuration is a topological Weyl semimetal (red curve). This topological phase is destroyed by 6% $Mn_{Bi}$, leading to a trivial insulating state (black curve).

We also calculate other defect concentrations to learn the evolution of electronic structure as defect density varies. Fig. 4C (red curve) summarizes the DFT calculated Γ point energy gap as a function of defect density, showing a gap closing at around 2% $Mn_{Bi}$ at which a topological phase transition from a Weyl semimetal to a trivial insulator occurs and illustrating the sensitivity of the topological phase of ferromagnetic bulk $MnBi_2Te_4$ to $Mn_{Bi}$ concentration. Following a procedure similar to that employed in Ref.22 key model parameters, such as the exchange potential strength $m_F$ which is equal to sums and differences of $J_S$ and $J_D$ in the ferromagnetic and antiferromagnetic configurations, can be readily extracted from parallel-spin-configuration DFT energy bands. Fig. 4C (blue curve) plots the defect density dependence of $m_F$, showing that it decreases nearly linearly with an increase of defect density. In Fig. 4C we normalize $m_F$ relative to the same-layer and adjacent-layer Dirac cone hybridization parameters $\Delta_S$ and $\Delta_D$. (A detailed explanation of these parameters can be found in Supplementary Fig. S6.)

Using the parameters extracted from DFT, we performed coupled Dirac cone model calculations for thin films with antiferromagnetic spin-configurations. Fig. 4D summarizes the calculated Γ point energy gaps for both spin-aligned bulk and antiferromagnetic thin film cases. The bulk results (dashed black line) agree well with the DFT results (gold diamonds), confirming the validity of our model. For antiferromagnetic thin films such as 4 and 6 SLs, gaps can be suppressed by half with less than 6% $Mn_{Bi}$. As is expected [44] (see supplementary Fig. S8 for detailed discussion), this effect is even more dramatic for odd SLs, as exemplified by the 3 and 5 SLs results in Fig. 4D. In that case, gap values show a linear dependence on the defect density, and a topological phase transition occurs at around 1% for 3 SL and 6% for 5 SL thin films. A turning point occurs at about 9% due to the exchange coupling contributed by the $Mn_{Bi}$ overtaking that by the central Mn layer. This eventually leads to a second gap crossing zero accompanied by a trivial insulator to Chern insulator transition with the opposite relationship between Chern number and magnetization. The gap quenching defect concentration (1%) is small for 3 SL $MnBi_2Te_4$ in our model calculations because of accidental proximity to a topological phase transition (22). However, the experimental Dirac mass gap for 3 SL is larger than that in the model calculation. Thus, its gap quenching should occur at a higher defect concentration.

This model captures the variation of the Dirac mass gap as the magnetic antisite density increases but neglects the antisites-induced disorder effect. In practice, observation of the QAH effect will require even lower $Mn_{Bi}$ antisite densities than those predicted here. Current bulk $MnBi_2Te_4$ has a 3% to 6% $Mn_{Bi}$ (26, 33, 39, 45) (10% to 20% of $Mn_{Sb}$ for $MnSb_2Te_4$ (31, 32)) concentration. A large density of $Bi_{Mn}$ antisites can also be inferred from the heavily degenerate n-type nature of the material. Considering the disorder effect, it is understandable why the QAH effect is not universally observed in an odd-SL $MnBi_2Te_4$ thin-film (16, 17, 19, 20, 46).

Our experimental STM/S study of intrinsic MBE-grown $MnBi_2Te_4$ thin films provides unambiguous evidence for a Dirac mass gap in regions with low magnetic antisite defect density and resolves the current debate as to whether or not the topological surface states are gapped. STS spatial mapping across boundaries between pristine and defective regions directly reveals how $Mn_{Bi}$ defects suppress the Dirac mass gap. *Ab initio* DFT and coupled Dirac cone model calculations unveil the microscopic mechanism for this correlation. The model calculations predict critical antisite densities above which the QAH effect cannot be observed in the antiferromagnetic $MnBi_2Te_4$ thin films (~6% for 5 SL films). By demonstrating the microscopic origin of the Dirac mass gap variation in $MnBi_2Te_4$, our work establishes magnetic defect as an important parameter in controlling the topological quantum phases.

**Materials and Methods**

Sample growth and STM/STS measurements

$MnBi_2Te_4$ thin films were grown in a home-built MBE chamber with base pressure at ~ $10^{-10}$ torr. High-oriented pyrolytic graphite (HOPG) substrates were cleaved in air and immediately transferred into the MBE chamber. HOPG substrates were outgassed at ~ 300 °C overnight before the growth of $MnBi_2Te_4$. High-purity Mn (99.99%), Bi (99.999%), and Te (99.999%) were evaporated from standard Knudsen cells. Samples were grown at 240 °C and post-annealed at the growth temperature in a Te ambient. Samples were transferred from the MBE chamber into the STM chamber, with base pressure ~ $10^{-11}$ torr, through a transfer vessel, with base pressure ~ $10^{-10}$ torr, to maintain the cleanness of the film. STM/STS measurements were conducted at 4.3 K. The W tip was prepared by electrochemical etching, and then cleaned by *in situ* electron-beam heating. STM d$I$/d$V$ spectra were measured using a standard lock-in technique with feedback loop off, whose modulation frequency is 490 Hz.

*Ab initio* DFT calculations

DFT calculations were performed using Vienna *Ab initio* Simulation Package (VASP) (47) in which generalized gradient approximations (GGA) of Perdew-Burke-Ernzerhof (PBE) (48) have been adopted for exchange-correlation potential. On-site correlation on the Mn-3d states is treated by performing a simplified (rotationally invariant) approach to the LSDA+U calculations (49) with $U - J$ equals 5.34 eV. U and J represents the strength of the effective on-site Coulomb and exchange interactions. A 20 Å of vacuum is used when calculating the $MnBi_2Te_4$ thin films to avoid the periodic image interactions normal to the surface. The global break condition for the electronic self-consistency loop is set to be $10^{-7}$ eV, and the plane wave energy cutoff is set to be 600 eV. DOS in Fig. 2E are calculated on a 16 × 16 × 1 Γ-centered k-point integration grid with a Gaussian broadening factor of 50 meV.

Supercells that correspond to 3 × 3 × 1 and 4 × 4 × 1 of the original primitive cell were used to simulate the antisite defects in the ferromagnetic configuration. As illustrated in Fig. S7, one $Mn_{Bi}$ antisite per sixteen Bi atoms in a 4 × 4 × 1 supercell corresponds to $Mn_{Bi}$ density of 1/16; one $Mn_{Bi}$ antisite per nine Bi atoms in a 3 × 3 × 1 supercell corresponds to $Mn_{Bi}$ density of 1/9; two $Mn_{Bi}$ antisites per sixteen atoms in a 4 × 4 × 1 supercell corresponds to $Mn_{Bi}$ density of 1/8. The global break condition for the electronic self-consistency loop is set to be $10^{-6}$ eV, and a 9 × 9 × 3 Γ-centered k-point integration grid was employed.

**Acknowledgments**

This work was primarily supported by the National Science Foundation (NSF) through the Center for Dynamics and Control of Materials: an NSF MRSEC under cooperative agreement no. DMR-1720595 and the US Air Force grant no. FA2386-21-1-4061. Other supports were from NSF grant nos. DMR-1808751, the Welch Foundation F-1672. Work at ORNL was supported by the U.S. Department of Energy, Office of Science, Basic Energy Sciences, Materials Sciences and Engineering Division. We thank C.Z. Chang for the helpful discussion in MBE growth.

**Author Contributions**

M.L. and C.-K.S. designed the research. M.L., Y.L., and L.F. performed the STM/STS study. H.K. and M.L. carried out the MBE growth. C.L. performed the theoretical calculation. J.Y. provided bulk $MnBi_2Te_4$ crystals. M.L., C.L., A.H.M., and C.-K.S. analyzed the data. M.L., A.H.M., and C.-K.S. wrote the paper with input from all the authors.

**Competing Interest Statement**

The authors declare no competing interest.

# Figures

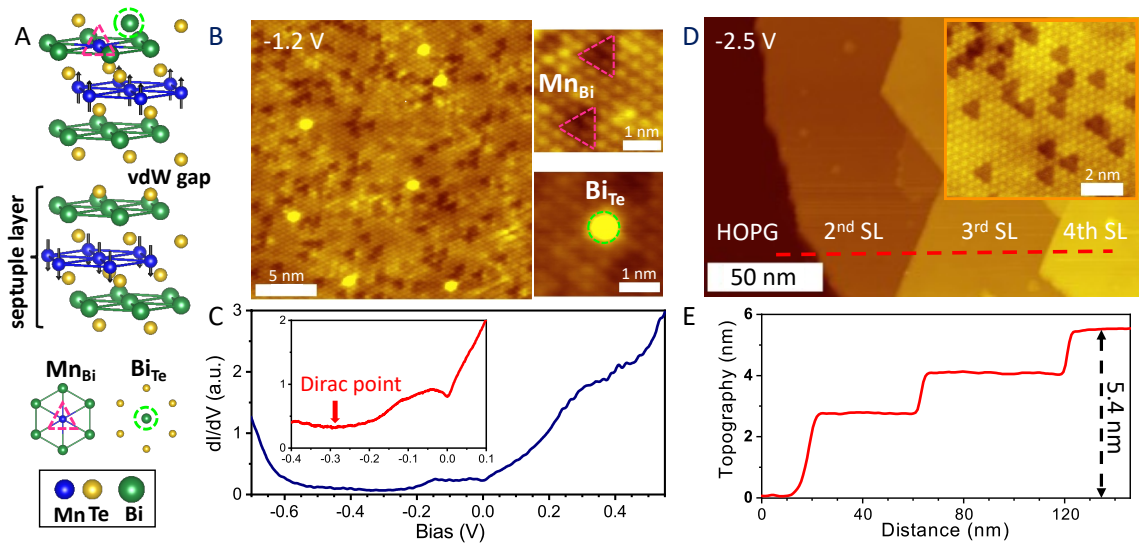

**Fig. 1. Bulk crystals versus MBE grown thin films.** (*A*), Crystal structure of MnBi$_2$Te$_4$. The black arrows illustrate the intralayer ferromagnetic and interlayer antiferromagnetic coupling. Cleave plane happens at the van der Waals gap. Mn$_{Bi}$ and Bi$_{Te}$ antisite defects are outlined with a pink triangle, and green circle. (*B*), Atomic image taken on Te truncated bulk MnBi$_2$Te$_4$ surface. Examples of Mn$_{Bi}$ and Bi$_{Te}$ antisites are outlined by pink triangles and green circles. (*C*), STS taken on bulk MnBi$_2$Te$_4$ surface. Inset red curve shows zoom-in bias range. The red arrow marks the Dirac point. (*D*), Topographic image taken on MBE-grown MnBi$_2$Te$_4$. Inset shows its atomic image. (*E*), Topographic cross-section along the red dashed line on D shows the thickness of each SL is 1.35 nm.

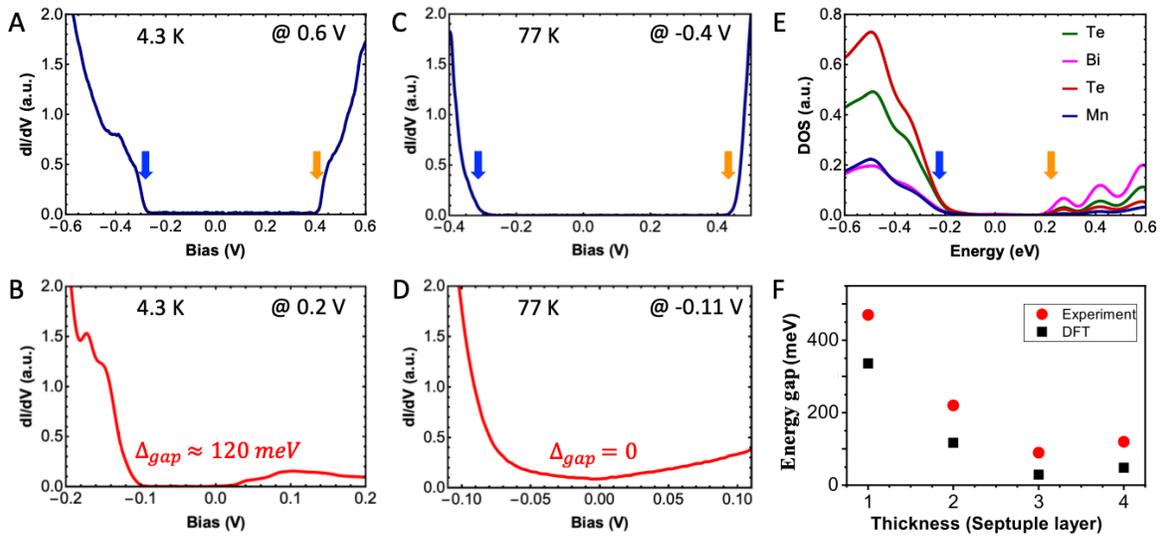

**Fig. 2. Direct observation of Dirac mass gap.** (*A,B*), STS taken on 4 SL MnBi$_2$Te$_4$ at 4.3 K with a 0.6 V and 0.2 V setpoint biases. STS in B shows a ~120 $meV$ Dirac mass gap. (*C,D*), STS taken on 4 SL MnBi$_2$Te$_4$ at 77 K with a -0.4 V and -0.11 V setpoint biases. STS in D shows that the Dirac mass gap has disappeared. (*E*), DFT calculated DOS for 4 SL MnBi$_2$Te$_4$ with a projection to the topmost four atomic layers. Blue and orange arrows on *A*, *C*, and *E* mark the threshold energies that bound the low-conductance bias region. (*F*), Summarized STS-measured and DFT-calculated energy gap as a function of film thickness.

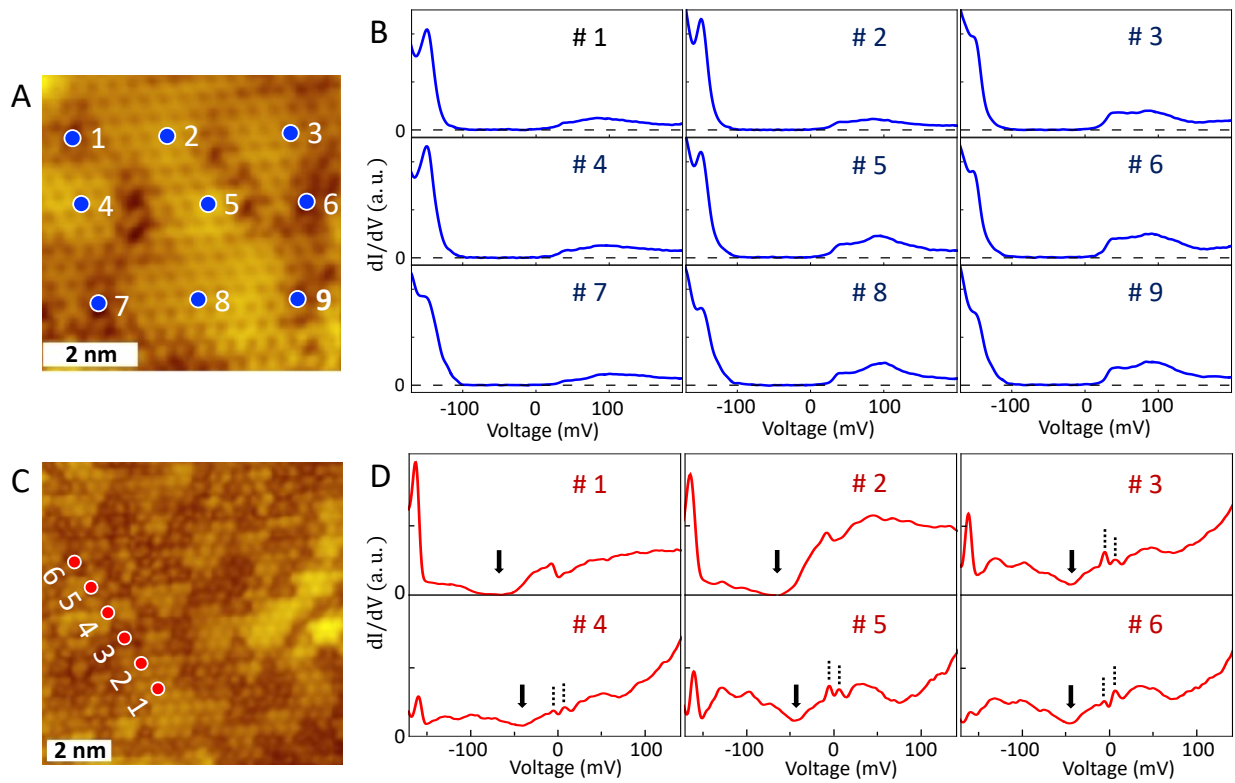

**Fig. 3. Dirac mass gap variation as a function of defect distribution.** (*A*), Atomic image taken on low defect 4 SL region (setpoint bias: $0.3\ V$). (*B*), Spatial-dependent STS distribution with spatial locations marked on *A*. Horizontal black dashed lines mark zero. (*C*), Atomic image taken on high defect 3 SL region (setpoint bias: $-0.35\ V$). (*D*), Spatial-dependent STS distribution with spatial locations marked on *C*. Black arrows mark the Dirac point. Vertical black dashed lines on STS #3 to #6 mark the zero-bias tunneling anomaly.

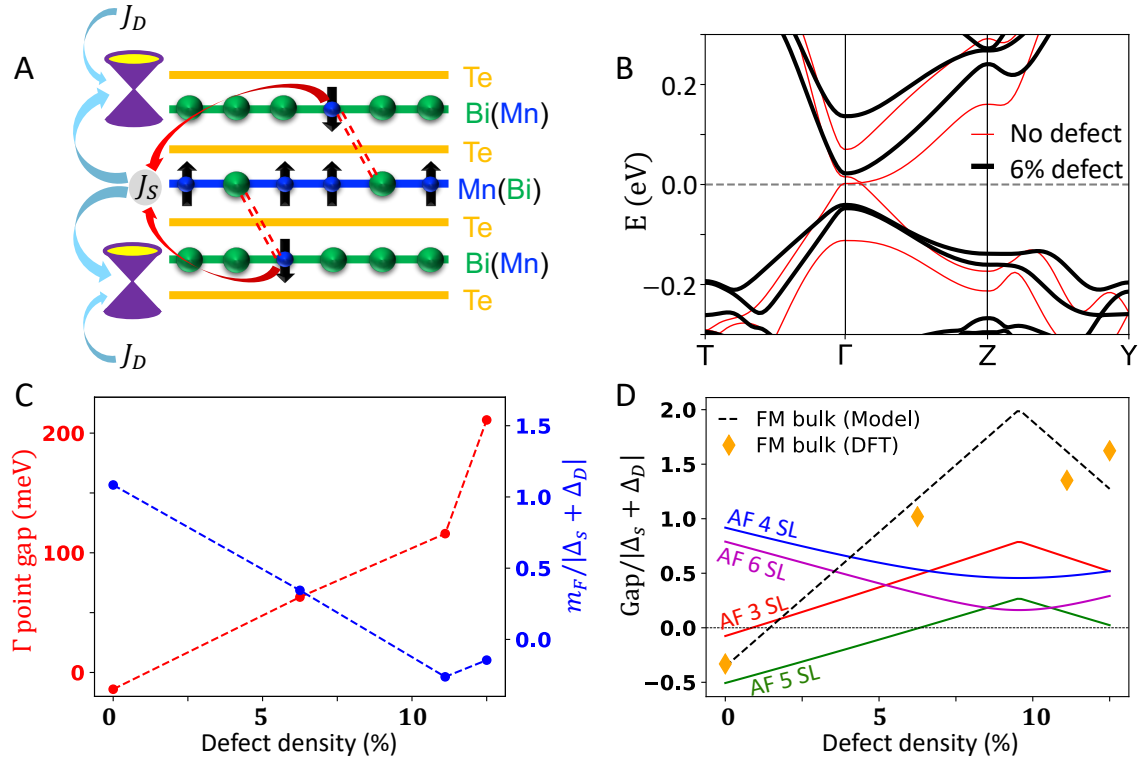

**Fig. 4. *Ab initio* DFT calculation and coupled Dirac cone model.** (*A*), Schematics for magnetic moments configuration and exchange couplings between the Dirac cones and the local magnetic moments. Both the $Mn_{Bi}$ antisites and Mn layer contribute to the exchange coupling. Red dashed lines illustrate the next nearest-neighboring antisites pairs. (*B*), Calculated band dispersion for ferromagnetic bulk $MnBi_2Te_4$ in defect-free and 6% $Mn_{Bi}$ antisites cases. (*C*), DFT calculated Γ point energy gap and extracted model parameter $m_F$ as a function of $Mn_{Bi}$ antisites defect density. Negative gap values represent inverted bands. (*D*), Coupled Dirac cone model-calculated Γ point energy gap as a function of $Mn_{Bi}$ antisites defect density in the case of ferromagnetic (FM) bulk and antiferromagnetic (AF) thin films. DFT calculated FM bulk results are plotted in orange diamonds. FM (AF) refers to spin moments in Mn layers being aligned (anti-aligned) between adjacent septuple layers.

# Supplementary Information for

# Visualizing the interplay of Dirac mass gap and magnetism at nanoscale in intrinsic magnetic topological insulators

Mengke Liu[1], Chao Lei[1], Hyunsue Kim[1], Yanxing Li[1], Lisa Frammolino[1], Jiaqiang Yan[2], Allan H. Macdonald[1]*, Chih-Kang Shih[1]*

\* To whom correspondence may be addressed.
**Email:** macdpc@physics.utexas.edu or shih@physics.utexas.edu.

**This PDF file includes:**

    Figures S1 to S8

    Details on coupled Dirac cone model and DFT supercells

    Additional discussion on magnetic antisites effect in odd and even septuple layers

**Fig. S1**

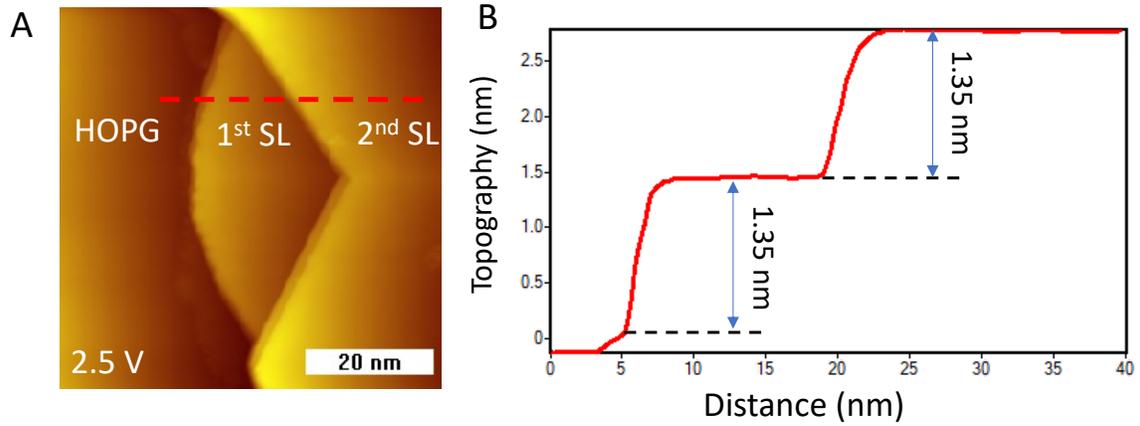

**Fig. S1. Additional data on MBE grown MBT on HOPG substrate.** (*A*), Topographic image taken on MBE-grown MBT showing the 1st and 2nd septuple layers (SLs). (*B*), Topographic cross section along the red dashed line on **a** showing the step height of each septuple layer is 1.35 nm.

**Fig. S2**

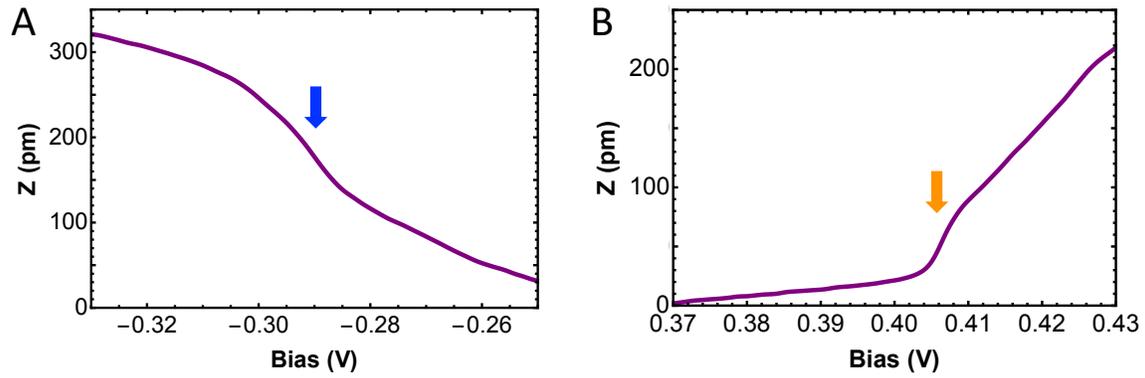

**Fig. S2. Characterization of tip-to-sample distance (Z) behavior at different biases.** (*A,B*), constant-current Z-V spectra at negative and positive biases show a Z jump at energies marked by blue and orange arrows. This Z-jump indicates a reduced tip-to-sample distance due to a reduced conductance. These two critical energies here coincide with the two thresholds in *dI/dV* spectra shown in Fig. 2A, confirming the reduced density-of-state bias region as discussed in the main text.

**Fig. S3**

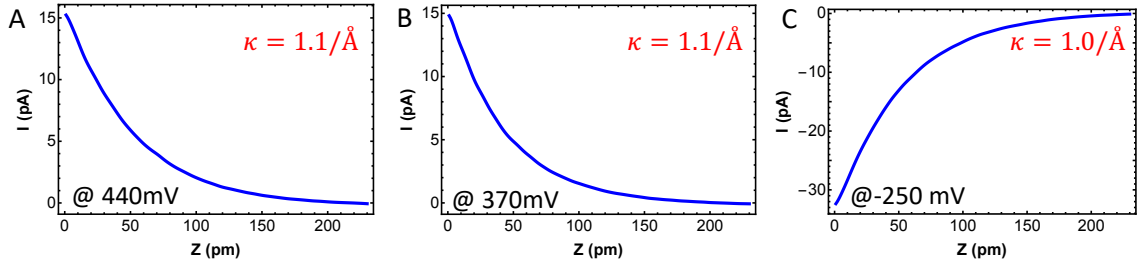

**Fig. S3. Characterization of tunneling barrier condition at different setpoint biases.** (*A-C*), *I-Z* spectra demonstrating the decay constant of the tunneling barrier at different setpoint biases. *I* denotes the tunneling current. *Z* denotes the tip-to-sample distance. The measured decay constant $\kappa$ from these *I-Z* spectra is 1.1/Å at 440 mV and 370 mV, 1.0/Å at -250 mV, indicating normal vacuum tunneling barriers above and below the threshold energies. This set of tunneling barrier characterization ensures that the *dI/dV* spectra in Fig. 2 are carried out in vacuum tunneling conditions without tip-to-sample interactions, thus confirming the reliability of the *dI/dV* spectra.

**Fig. S4**

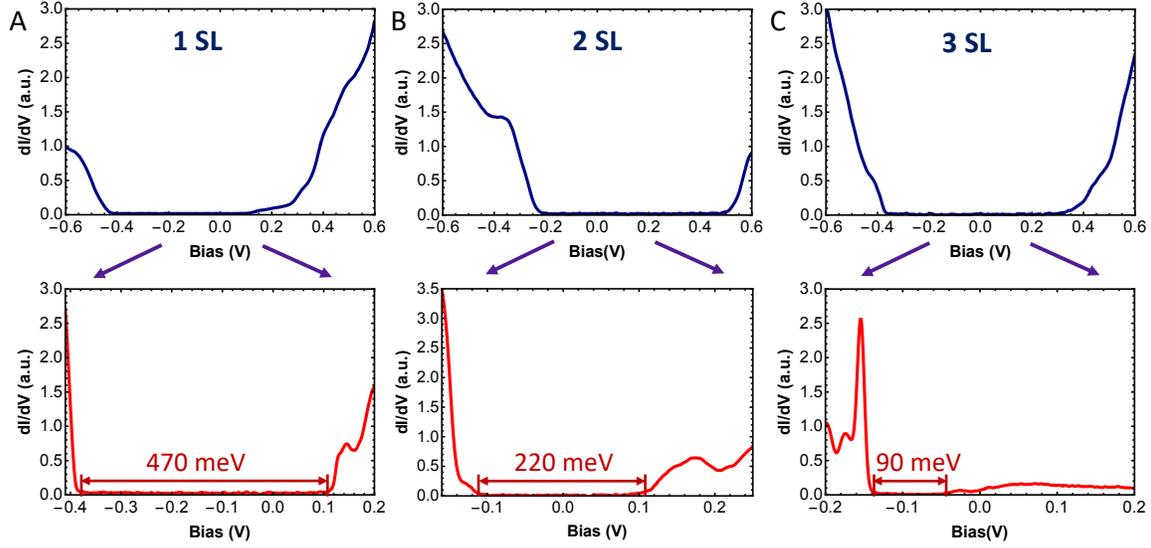

**Fig. S4. STS on different septuple layer MBT showing thickness-dependent energy gap.** (*A*), STS taken on 1 SL showing 470 meV gap. (*B*), STS taken on 2 SL showing 220 meV gap. (*C*), STS taken on 3 SL showing 90 meV gap.

**Fig. S5**

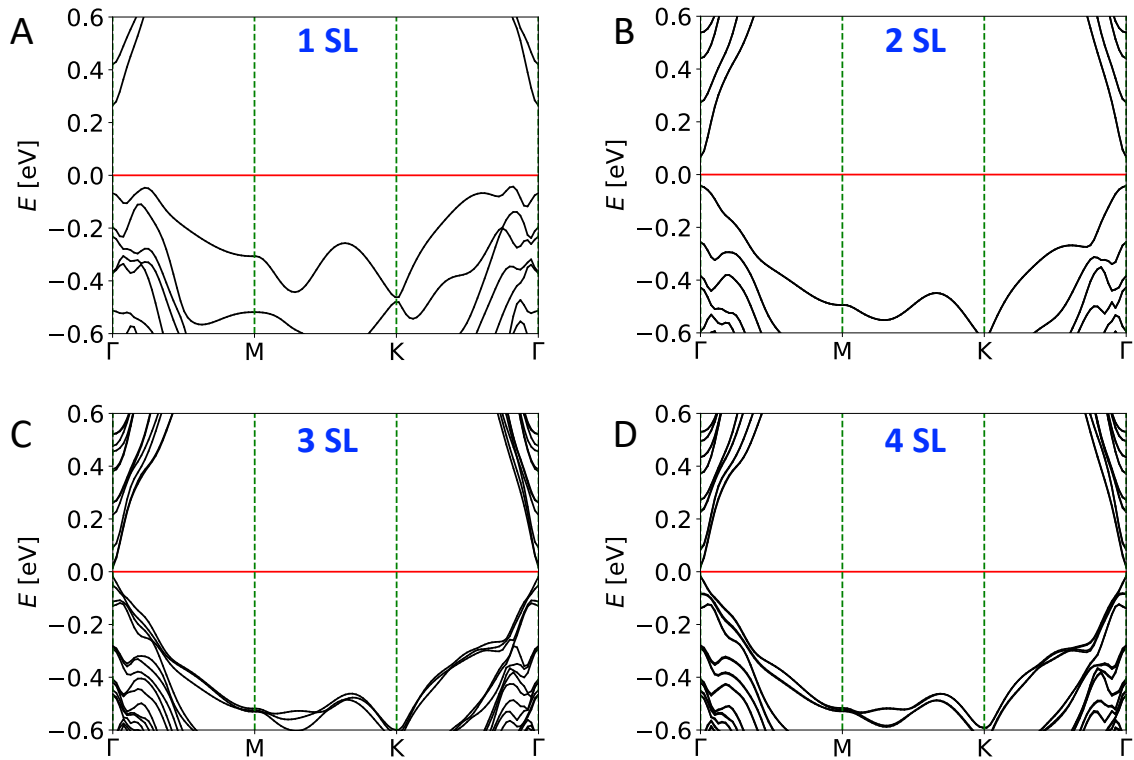

**Fig. S5. DFT calculated MBT thin film band structures.** (*A-D*), DFT calculated 1 SL to 4 SL MBT band structure. The calculated energy gap is: 336 meV (1SL), 117 meV (2 SL), 29 meV (3 SL), 48 meV (4 SL).

**Details on Coupled Dirac cone model and DFT supercells**

We employed the coupled Dirac cone model (Ref. 22 in main text) that includes only Dirac cone surface states on the top and bottom surfaces of each septuple layer and hoppings between Dirac cones.

The Hamiltonian is expressed as:

$$H = \sum_{k_\perp,ij}\left[\left((-)^i \hbar v_D (\hat{z}\times\sigma)\cdot k_\perp + m_i\sigma_z\right)\delta_{ij} + \Delta_{ij}(1-\delta_{ij})\right]c^\dagger_{k_\perp i}c_{k_\perp j},$$

where $i$ and $j$ are Dirac cone labels, $\hbar$ the reduced Planck constant, and $v_D$ the velocity of the Dirac cones. The parameter $m_i$ is the exchange potential strength of the $i$th Dirac cone. Hopping from the $i$th surface to the $j$th surface is denoted by $\Delta_{ij}$. Here we only retain the most important hybridization parameters indicated by the hopping within the same layer $\Delta_S$, and the hopping across the van der Waals gap between adjacent layers $\Delta_D$; The total exchange potential in bulk MnBi$_2$Te$_4$ for each Dirac cone is $m_F = J_S + J_D$ in the ferromagnetic configuration, and $m_{AF} = J_S - J_D$ in the antiferromagnetic configuration, where $J_S$ and $J_D$ are the exchange couplings from the same layer and the adjacent layer, as illustrated in Fig. S6. The existence of the Mn$_{Bi}$ antisites defects, forming an antiferromagnetic coupling with that in the central Mn layer[23], leading to a reduced same-septuple-layer and neighboring-septuple-layer exchange couplings $J_S$ and $J_D$. The presence of Bi$_{Mn}$ antisite also decreases $J_S$ by reducing the magnetic moments in the Mn layer.

For the ferromagnetic configuration, the exchange energies are the same in every layer, and the model band energy dispersion along the $\Gamma$ to Z line is:

$$E(k_z) = \pm\sqrt{\Delta_S^2 + \Delta_D^2 + 2\Delta_S\Delta_D \cos k_z d} \pm m_F.$$

The energies at $k_z d = 0$ and $k_z d = \pi$ are:

$$E_\Gamma = \pm(\Delta_S + \Delta_D) \pm m_F;\ E_Z = \pm(\Delta_S - \Delta_D) \pm m_F.$$

Comparing with DFT band energies at Γ and Z points, one can extract the model parameters, $\Delta_S$, $\Delta_D$, and $m_F$.

To estimate the exchange couplings as a function of defect density, $J_S(x)$ and $J_D(x)$ where $x$ denotes the Mn$_{Bi}$ defect density, we performed DFT calculations at a few different Mn$_{Bi}$ antisites densities, 0%, 6%, 11%, 12.5%, as explained in the Methods and Fig. S7. Assuming the ratio between $J_D$ and $J_S$ is independent of the defect density, i.e. $J_D(x) = \delta J_S(x)$, $\delta$ turns out to be 0.8, extracted from the DFT calculated results. Fig. 4 demonstrates the exchange coupling $m_F$ decreases nearly linearly with defect density. It is thus reasonable to assume $J_S(x) = J_S^0 - \alpha x$, where $J_S^0$ is the exchange coupling without defect and $\alpha$ is the linear coefficient. Therefore, the effective exchange potential in bulk MnBi$_2$Te$_4$ with ferromagnetic spin configuration can be expressed as:

$$m_F(x) = (1 + \delta)(J_S^0 - \alpha x).$$

We may then extract the value of $\alpha$ from Fig. 3C, a summarized $m_F$ as a function of defect density.

These exchange coupling dependence on the defect density $J_S(x)$ and $J_D(x)$, are then used to perform model calculations for the ferromagnetic bulk and antiferromagnetic thin films, as shown in the main text Fig. 4D.

**Fig. S6**

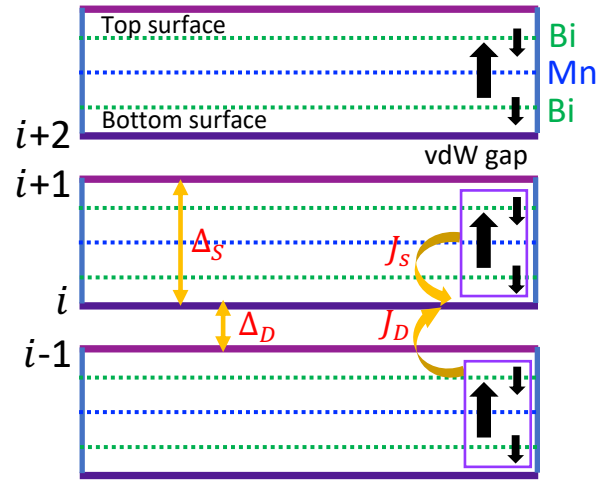

**Fig. S6. Schematic illustration of coupled Dirac cone model in the ferromagnetic configuration.** Dirac cones localized on each septuple layer's top and bottom surfaces are hybridized with remote Dirac cones. $\Delta_S$ denotes hopping parameter between Dirac cones from the same septuple layer. $\Delta_D$ denotes hopping parameter between Dirac cones from the adjacent layer across the van der Waals gap. Dirac cones are exchange-coupled to local magnetic moments from their own layer ($J_S$) and their neighboring layer ($J_D$). Black arrows represent the local orientation of the magnetic moments. $i$ denotes the layer index. "Ferromagnetic configuration" refers to spin moments in Mn layers being aligned between adjacent septuple layers.

**Fig. S7**

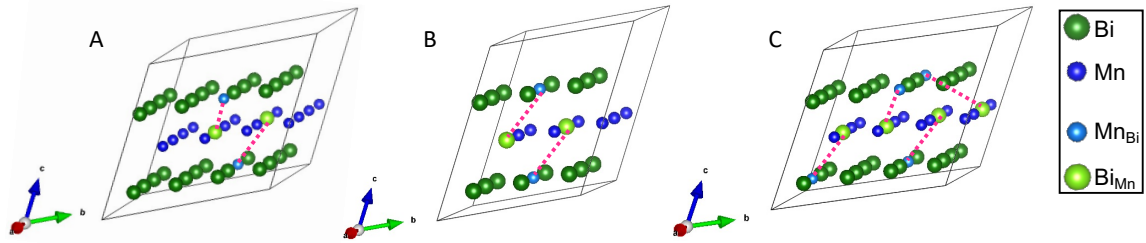

**Fig. S7. DFT supercell configurations for bulk MnBi$_2$Te$_4$ with antisite defects.** (*A*), A $4 \times 4 \times 1$ supercell configuration corresponding to 6% Mn$_{Bi}$ antisite density. one per sixteen Bi atoms is substituted by its next-nearest neighboring Mn atom. (*B*), A $3 \times 3 \times 1$ supercell configuration corresponding to 11% Mn$_{Bi}$ antisite density. One per nine Bi atoms is substituted by its next-nearest neighboring Mn atom. (*C*), A $4 \times 4 \times 1$ supercell configuration corresponding to 12.5% Mn$_{Bi}$ antisite density. Two per sixteen Bi atoms were substituted by their next-nearest neighboring Mn atoms. The red dashed lines connect the next-nearest neighboring Mn$_{Bi}$ and Bi$_{Mn}$ antisites pairs. The antisite defect density in the top and bottom Bi layers are kept the same. Te atoms are left out for simplicity.

**Additional discussion on magnetic antisites effect in odd and even septuple layers**

Fig. 4D in the main text shows that for odd septuple layers, the Dirac mass gap exhibits a nearly linear suppression as a function of defect density, and a negative gap eventually collapses accompanied by a topological phase transition. However, for even septuple layers, the gap change as a function of defect density at small defect density appears linear and, when close to the minimum (9%), it appears like a quadratic curve. Their gap signs always remain positive, indicating no topological phase transition. Regardless of the gap sign, the absolute value of a gap appears to be more robust against magnetic antisite density in even septuple layers than that in odd septuple layers.

The behavior is under expectation from a toy model that has been discussed by Lei and MacDonald (ref. 44 in main text). Here we recap it below:

Antiferromagnetic $MnBi_2Te_4$ thin films can be simply illustrated with a model that only considers the two Dirac cones located on the top and bottom surfaces. The two Dirac cones have the same (opposite) sign of mass terms for MBT thin films with an odd (even) number of SLs. For thin films with odd SLs, the Hamiltonian may be written as:

$$H_{odd} = \begin{pmatrix} m & v_D k^+ & \Delta & 0 \\ v_D k^- & -m & 0 & \Delta \\ \Delta & 0 & m & -v_D k^+ \\ 0 & \Delta & -v_D k^- & -m \end{pmatrix},$$

where $k^{\pm} = k_x \pm i k_y$, $v_D$ is the Dirac velocity, $m$ is the Dirac mass that depends on the exchange coupling, $\Delta$ is the hybridization parameter between the two Dirac cones. The eigenvalues are: $E = \pm\sqrt{v_D^2 |k|^2 + (m \pm \Delta)^2}$, where $|k|^2 = k_x^2 + k_y^2$.

In this simplified model, the band gap locates at $\Gamma$ point, which is $E_{gap}^{odd} = 2(m - \Delta)$. We can see that this gap exhibits a linear dependence on the exchange mass $m$, and a topological phase transition appears at $m = \Delta$. Note that we assumed a linear dependence of $m$ on the antisites density, which has been above.

Similarly, for even SLs, the top and bottom Dirac cones have an opposite sign of Dirac mass. The Hamiltonian reads:

$$H_{even} = \begin{pmatrix} m & v_D k^+ & \Delta & 0 \\ v_D k^- & -m & 0 & \Delta \\ \Delta & 0 & -m & -v_D k^+ \\ 0 & \Delta & -v_D k^- & m \end{pmatrix},$$

It has eigenvalues as $E = \pm\sqrt{v_D^2 |k|^2 + m^2 + \Delta^2}$, and band gap as $E_{gap}^{even} = 2\sqrt{m^2 + \Delta^2}$. We see that in this case the gap is always positive indicating the disappearance of the topological phase transition. The gap minimizes at $m = 0$, where the exchange splitting contributed by antisites fully cancels that from the central Mn layer. A Taylor expansion around this local minimum shows quadratic behavior. Far away from this local minimum where $|m| > \Delta$, such a gap appears to be linearly dependent on $m$.

Fig. S8 plots band structure evolution as a function of exchange mass $m$ to visualize the energy gap change as a function of $m$ in odd and even SLs, demonstrating the topological phase transition and the sensitivity to exchange mass of the energy gap in odd SLs films.

**Fig. S8**

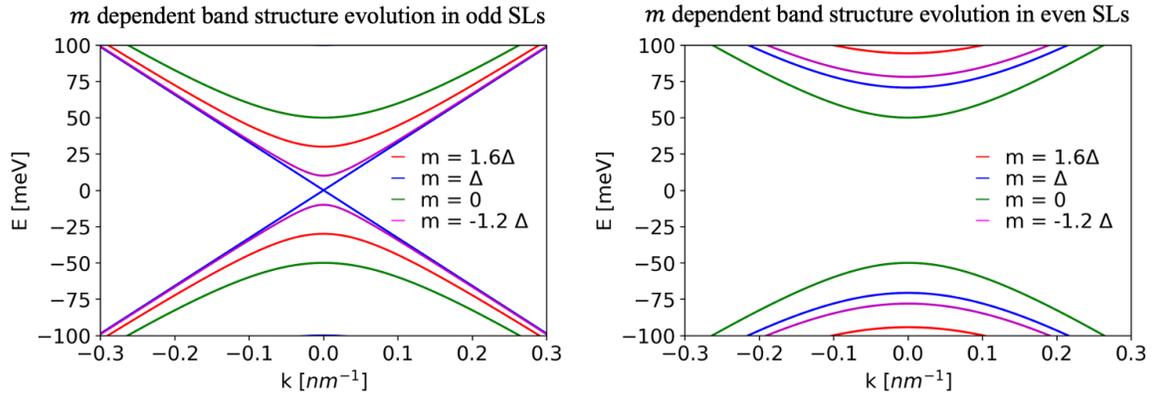

**Fig. S8. Band structure evolutions as a function of exchange mass $m$, in antiferromagnetic MBT thin films.** (Left) Odd septuple layer thickness. (Right) even septuple layer thickness. $m$ is parametrized in the magnitude of the hybridization parameter $\Delta$, and $\Delta$ is chosen to be 50 meV.